\documentclass[12pt]{article}
\usepackage{amssymb,graphicx,url}
\usepackage{epstopdf}
\usepackage{amsmath,amsfonts}
\usepackage{epsfig,color}

\begin{document}

\begin{flushright}  
INR-TH/2014-018
\end{flushright}
\vskip -0.9cm

\sloppy

\title{\bf Revisiting constraints on (pseudo)conformal Universe with Planck data}
\author{G.~I.~Rubtsov$^{a}$\footnote{{\bf e-mail}:
    grisha@ms2.inr.ac.ru} ~and 
S.~R.~Ramazanov$^{b}$\footnote{{\bf e-mail}: Sabir.Ramazanov@ulb.ac.be} 
\\
$^a$ \small{\em Institute for Nuclear Research of the Russian Academy of Sciences,}\\
\small{\em 60th October Anniversary st. 7a, Moscow 117312, Russia}\\
$^b$ \small{\em Universit\'e Libre de Bruxelles, Service de Physique Th\'eorique,}\\
\small{\em  CP225, Boulevard du Triomphe, B-1050 Brussels, Belgium}}

{\let\newpage\relax\maketitle}

\begin{abstract}
We revisit constraints on the (pseudo)conformal Universe from
the non-observation of statistical anisotropy in the Planck data. The
quadratic maximal likelihood estimator is applied to the Planck
temperature maps at frequencies 143 GHz and 217 GHz as well as their
cross-correlation. The strongest constraint is obtained in the
scenario of the (pseudo)conformal Universe with a long intermediate
evolution after conformal symmetry breaking. In terms of the
relevant parameter (coupling constant), 
the limit is $h^2 <0.0013$ at $95\%$ C.L. (using the
cross-estimator). The analogous limit is much weaker in the scenario without 
the intermediate stage ($h^2 \ln \frac{H_0}{\Lambda}<0.52$) allowing the 
coupling constant to be of order one. In the latter case, 
the non-Gaussianity 
in the 4-point function appears to be a more promising signature.

\end{abstract}

Statistical isotropy (SI) is one of the vanilla predictions of the
slow roll inflation. Deviations from this property -- if observed in
the cosmic microwave background (CMB) or large scale structure surveys
-- would imply a non-trivial extension of the standard
cosmology. Several models of inflation with vector fields have
been recently put forward~\cite{Ackerman:2007nb} (see
~\cite{Soda:2012zm} for a review) predicting an anisotropic Universe
with a detectable statistical anisotropy (SA). It was pointed out,
however, that many of these models suffer from
ghosts~\cite{Himmetoglu:2008hx} or rely on strong tuning in the
parameter space~\cite{Bartolo:2012sd}. These problems are absent in
some alternatives to inflation i.e., models of the (pseudo)conformal
Universe~\cite{Rubakov:2009np, Creminelli:2010ba,
  Hinterbichler:2011qk}. The latter are the main focus of this {\it Letter}.

In the pseudo(conformal) Universe, the space-time geometry is 
effectively Minkowskian at the times preceding the hot Big Bang. The state of the
early Universe in this picture is described in terms of conformal 
field theory. The conformal symmetry is assumed to be spontaneously
broken down to the de Sitter subgroup. The zero-weight conformal field
present in the Universe at these early times evolves in the 
symmetry breaking background and its perturbations
acquire flat power
spectrum~\cite{Rubakov:2009np, Creminelli:2010ba,
  Hinterbichler:2011qk}. These field perturbations get reprocessed
into adiabatic perturbations at much later epoch. 
The source of non-trivial phenomenology in
this setup is the interaction between zero-weight field perturbations
and the Goldstone field associated with the symmetry breaking
pattern~\cite{Libanov:2010nk, Libanov:2011bk,Hinterbichler:2012mv,
  Creminelli:2012qr}. In particular, very long wavelength modes of the
Goldstone field give rise to SA~\cite{Libanov:2010nk, Libanov:2011hh,
  Creminelli:2012qr}, while shorter ones lead to non-Gaussianity
(NG)~\cite{Libanov:2011bk, Creminelli:2012qr}.

Concrete realizations of the (pseudo)conformal Universe include
conformal rolling scenario~\cite{Rubakov:2009np} and Galilean
genesis~\cite{Creminelli:2010ba}. In these two models the conformal group
is spontaneously broken down to the de Sitter subgroup by the homogeneous
time-dependent solution of the unit conformal weight field $\rho$. The form of
the solution is fixed by the dilatation invariance, which remains
unbroken after spontaneous symmetry breaking,
\begin{equation}
\nonumber 
\rho =\frac{1}{h(t_* -t)} \; .
\end{equation}
The constant $h$ here is the most important parameter of the conformal rolling scenario 
and Galilean genesis, as it governs the non-trivial phenomenology, including SA; $t_*$ is the constant of integration, which 
has the meaning of the end-of-roll time. 

At the level of primordial curvature perturbations $\zeta$, SA implies 
directional dependence of the power spectrum, 
\begin{equation}
\label{powerspgen}
{\cal P}_{\zeta} ({\bf k}) \propto \left (1+ \sum_{LM} q_{LM} (k) Y_{LM} (\hat{{\bf k}}) \right) \; .
\end{equation}
Here $Y_{LM} (\hat{{\bf k}})$ are the spherical harmonics, $\hat{{\bf k}}$ is the 
direction of the perturbation wavevector ${\bf k}$ and $q_{LM} (k)$ are the coefficients parametrizing SA. 

In the (pseudo)conformal Universe scenario, 
there are two alternative predictions 
concerning SA. One of them is obtained if 
cosmological modes are superhorizon by the end of the
roll (at times close to $t_*$). In that case, the
directional dependence is of the quadrupolar form~\cite{Libanov:2010nk,
  Creminelli:2012qr},
\begin{equation}
\label{q2mrolling}
q_{2M} =\frac{H_0}{k}q'_{2M} +q''_{2M} \; .
\end{equation}
Here $q'_{2M}$'s and $q''_{2M}$'s encode contributions to SA appearing
in the linear and quadratic orders in the parameter $h$, respectively.
Note that the leading order (LO) contribution is characterized by the
decreasing amplitude; the Hubble rate $H_0$ plays the role of the
ultraviolet cutoff for infrared modes of the Goldstone field feeding
into SA. Coefficients $q'_{2M}$ obey the Gaussian statistics with
the zero mean and the following dispersion
 \begin{equation}
\nonumber 
\langle q'_{2M} q^{'*}_{2M'}\rangle = \frac{\pi h^2 }{25} \delta_{MM'} \; .
\end{equation}
The sub-leading order (SLO) contribution is of the axisymmetric form. Namely,
\begin{equation}
\nonumber 
q''_{2M}=-\frac{4\pi v^2}{5} Y^{*}_{2M} (\hat{{\bf v}})
\end{equation}
Here ${\bf v}$ is the Gaussian vector related to the Goldstone field. 
Its components have zero means and dispersions 
\begin{equation}
\nonumber 
\langle v^2_i \rangle =\frac{3h^2}{8\pi^2} \ln \frac{H_0}{\Lambda} \; ,
\end{equation}
where $\Lambda$ is the infrared cutoff for the modes of the Goldstone
field; $\hat{{\bf v}}={\bf v}/v$ is a unit vector.

 \begin{figure}[tb!]
\begin{center}
\includegraphics[width=0.32\columnwidth,angle=-90]{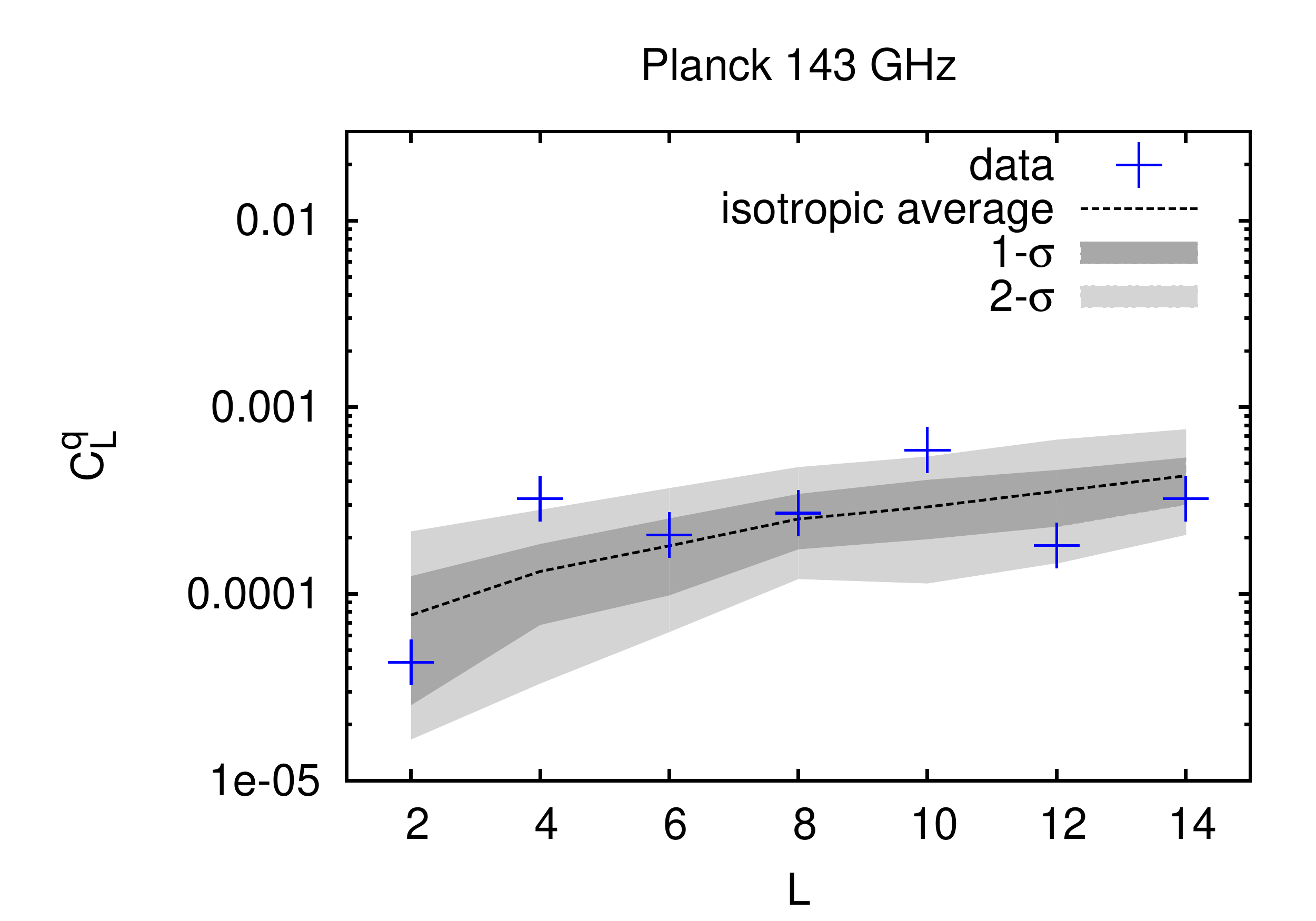}
\includegraphics[width=0.32\columnwidth,angle=-90]{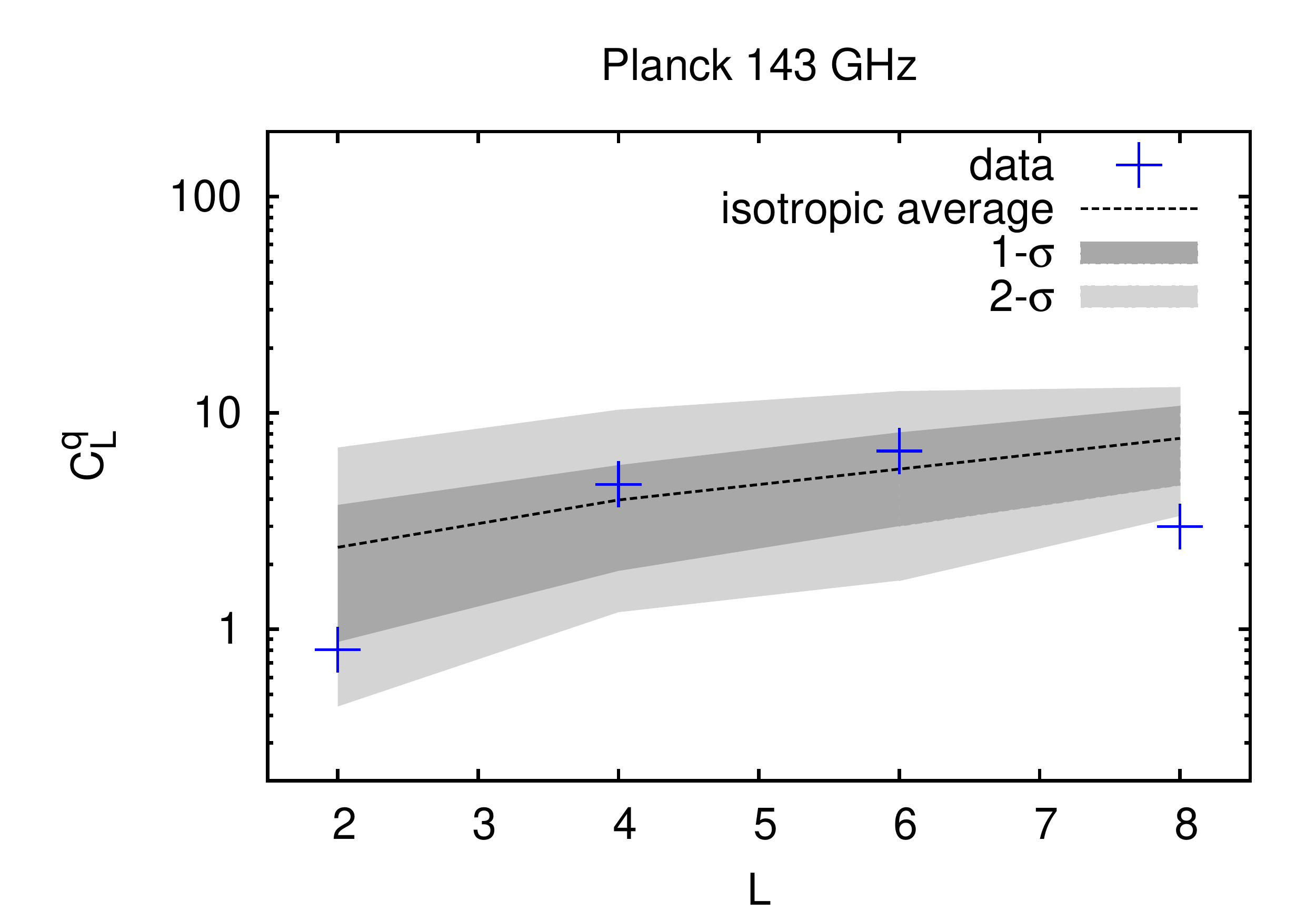}
\includegraphics[width=0.32\columnwidth,angle=-90]{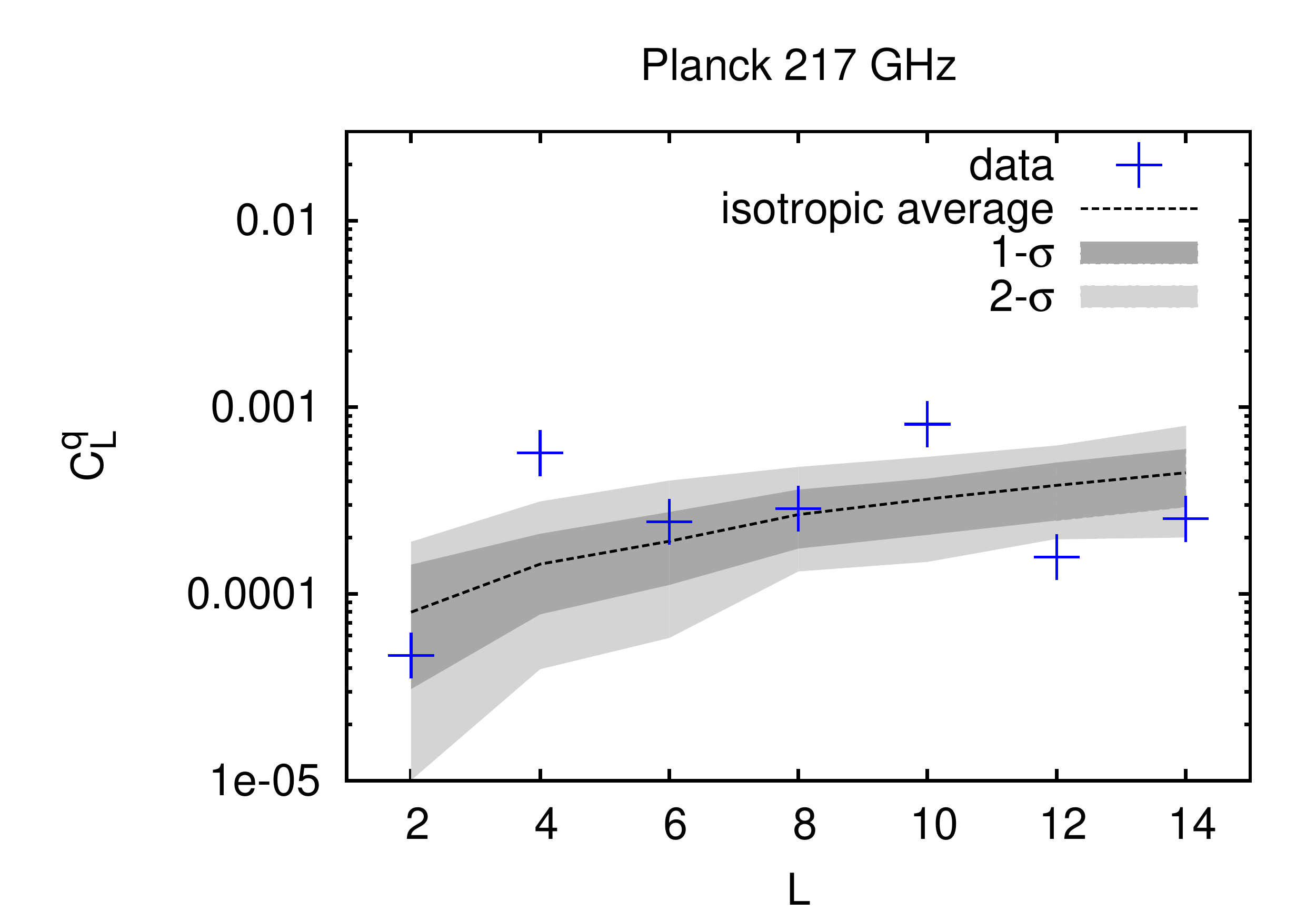}
\includegraphics[width=0.32\columnwidth,angle=-90]{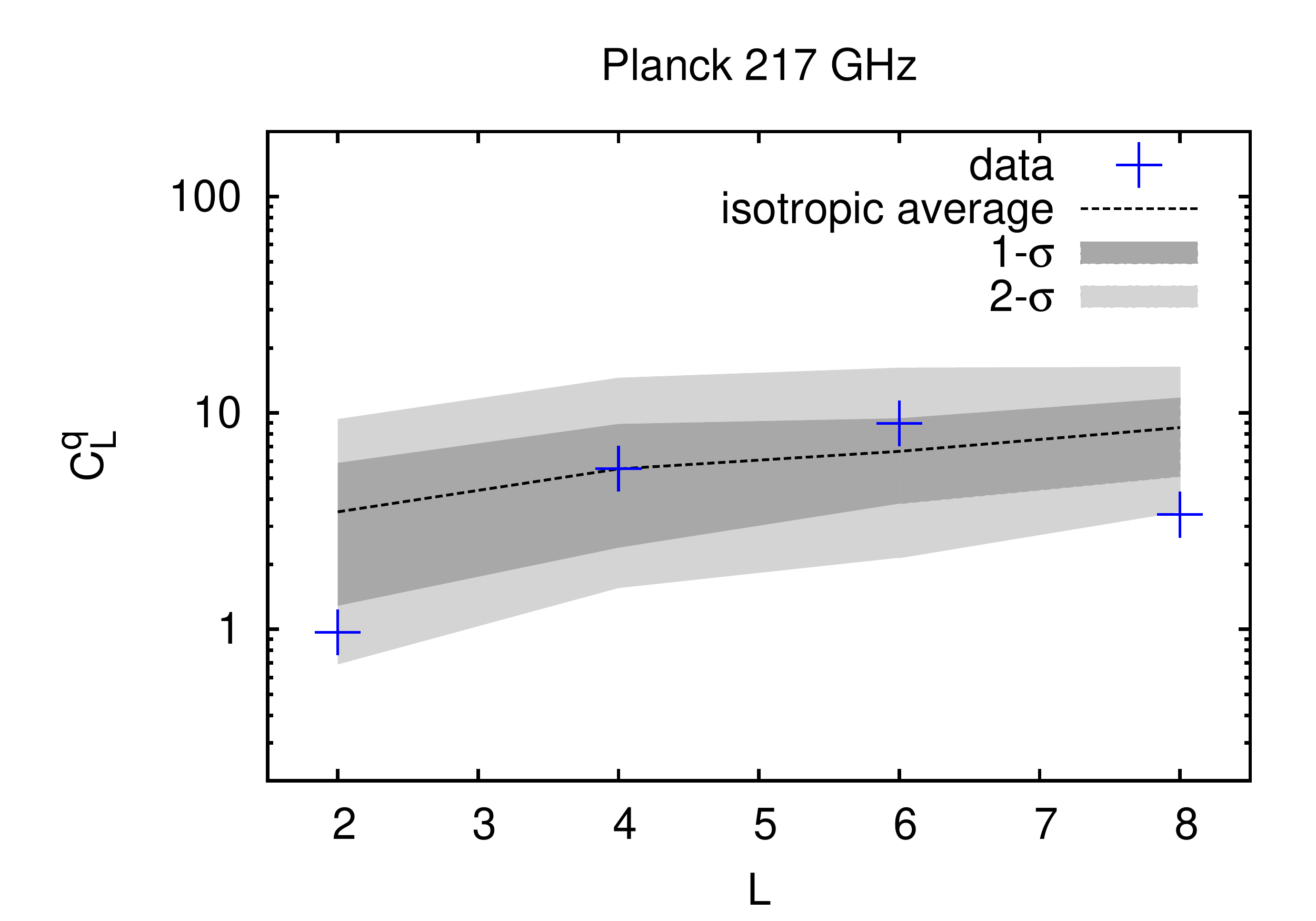}
\includegraphics[width=0.32\columnwidth,angle=-90]{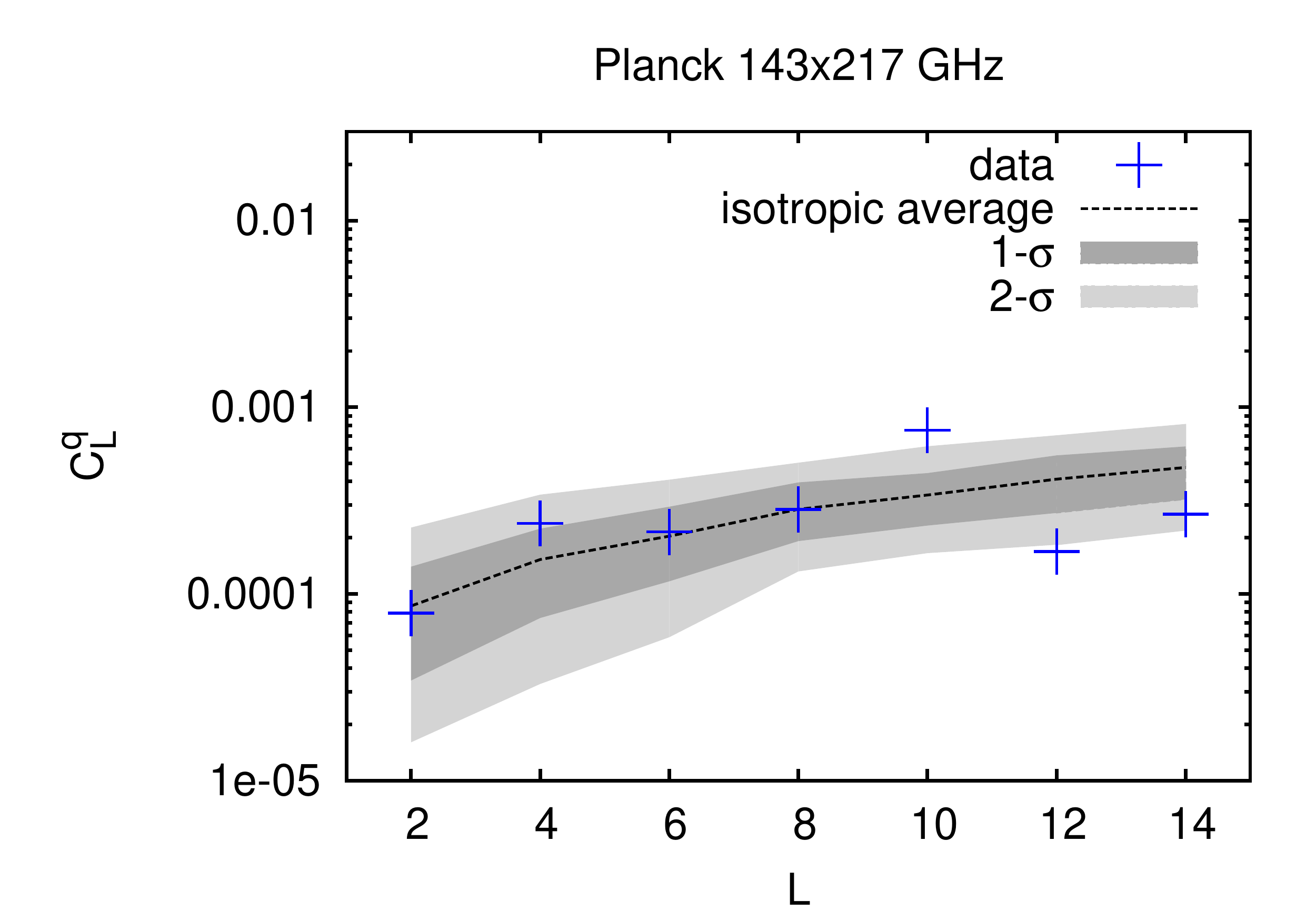}
\includegraphics[width=0.32\columnwidth,angle=-90]{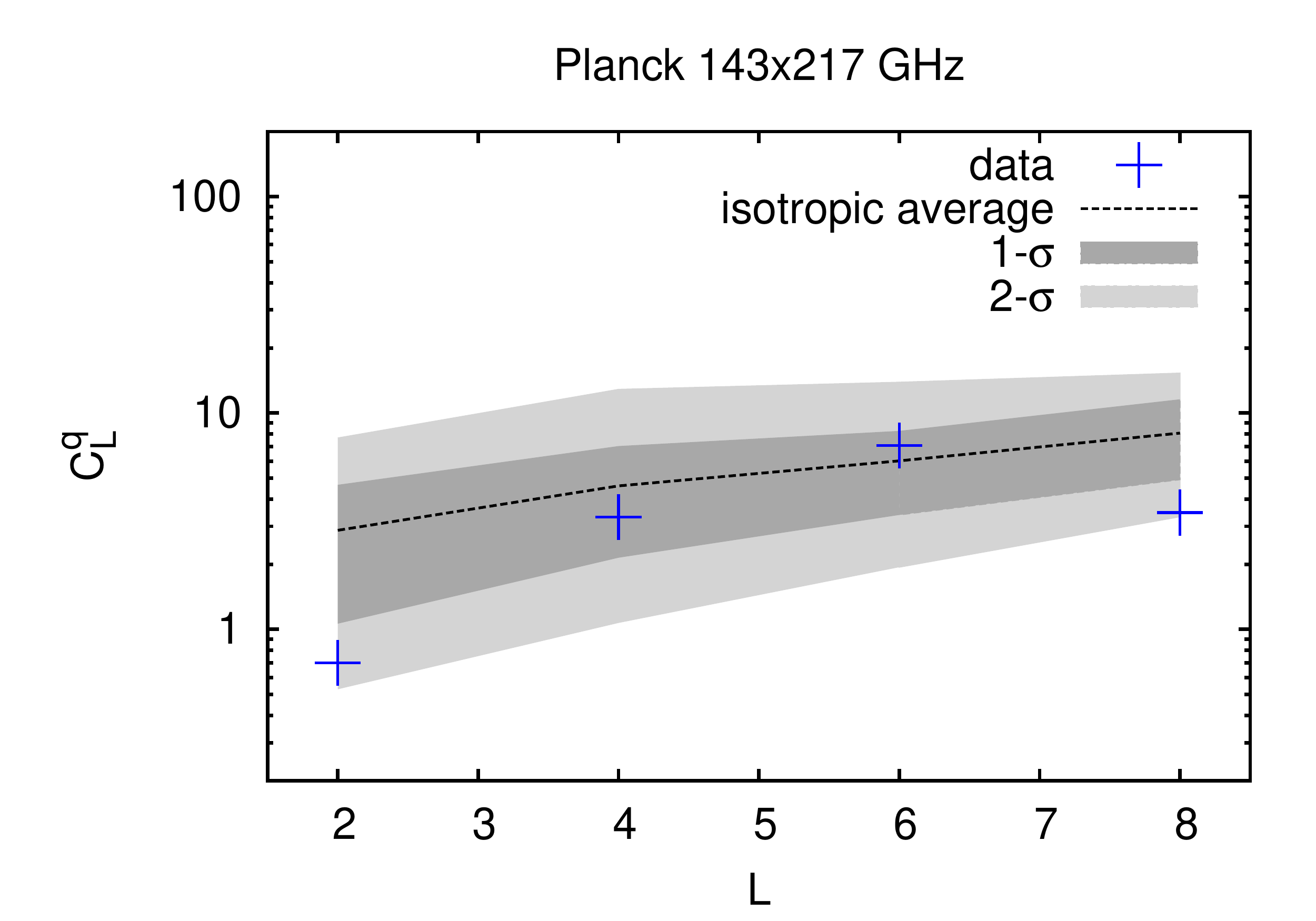}
\end{center}
\caption{Coefficients $C^{q}_L$ given by Eq.~\eqref{cql} reconstructed from the Planck data. Plots in the left and 
in the right column correspond to the choice $a(k)=1$ and $a(k)=H_0 k^{-1}$ in Eq.~\eqref{cll}, respectively. 68\% and 95\% C.L. intervals 
are overlaid with dark grey and light grey, respectively.\label{217crosscql}}
\end{figure}

Another prediction is obtained if cosmological modes of interest are still subhorizon by the end of
the roll. After the rolling stage, 
they proceed to evolve at the so called intermediate
stage~\cite{Libanov:2011hh, Mironov:2013bza}. The structure of SA in this case is
particularly rich. Namely, all the coefficients $q_{LM}$ with even $L$
are non-zero in~\eqref{powerspgen}. They are the Gaussian quantities with
zero means and variances given by~\cite{Libanov:2011hh}
\begin{equation}
\label{vr1} 
\langle q_{LM} q^{*}_{L'M'} \rangle =\tilde{Q}_L h^2\delta_{LL'} \delta_{MM'} \; ,
\end{equation}
where 
\begin{equation}
\nonumber 
\tilde{Q}_L =\frac{3}{\pi} \times \frac{1}{(L-1)(L+2)} \; .
\end{equation}
In what follows, we use the notion ``sub-scenario A''
 for the 
version of (pseudo)conformal Universe 
without the intermediate stage, Eq. \eqref{q2mrolling}, 
and ``sub-scenario B'' for the version with intermediate stage,
Eq.~\eqref{vr1}.

In order to derive the coefficients $q_{LM}$ from the CMB data, we
make use of the quadratic maximal likelihood (QML) estimator first
constructed in Ref.~\cite{Hanson:2009gu}. This has proved to be a
powerful tool for studies of SA in WMAP~\cite{Hanson:2009gu,
  Ramazanov:2012za, Ramazanov:2013wea} and Planck
data~\cite{kim}. Furthermore, QML methodology of the data analysis
results in an excellent agreement with the exact methods used in
Ref.~\cite{Groeneboom:2008fz}. The estimator is a quadratic
form in the space of maps. In this paper we rewrite the 
estimator as a bilinear function of two maps in a way similar 
to the WMAP cross-power spectrum~\cite{Hinshaw:2003ex}. Given the spherical harmonic
coefficients for the two maps $\hat{a}^{i}_{lm}$, $\hat{a}^{j}_{lm}$,
where $i$,$j$ denote the frequency band, we express both the original
estimator $i=j$ and the new cross-estimator $i\ne j$ in the following way
\begin{equation}
\label{qlm}
q^{ij}_{LM}=\sum_{L'M'} ({\bf F}^{ij})^{-1}_{LM;L'M'} (h^{ij}_{L'M'} -\langle h^{ij}_{L'M'} \rangle ) \; ,
\end{equation}
where $\langle \rangle$ denotes the averaging over different realizations of isotropic
maps and 
\begin{equation}
\label{hlm}
h^{ij}_{LM}=\sum_{ll'mm'} \frac{1}{2}i^{l'-l}C_{ll'} B^{LM}_{lm;l'm'} \bar{a}^{i}_{l,-m}  \bar{a}^{j}_{l'm'} \; ;
\end{equation}
$\bar{a}^{i}_{lm}$ are the CMB temperature coefficients 
filtered with the inverse isotropic covariance, 
\begin{equation}
\nonumber 
\bar{a}_{lm} =\left({\bf S}^{iso} +{\bf N}^{i} \right)^{-1}_{lm;l'm'} \hat{a}_{l'm'} \; .
\end{equation}
Here  
${\bf S}^{iso}$ is a theoretical isotropic covariance, and 
${\bf N}$ is the noise matrix. The coefficients $C_{ll'}$ in Eq.~\eqref{hlm} are given by
\begin{equation}
\label{cll}
C_{ll'}=4\pi  \int d \ln k \Delta_l (k) \Delta_{l'} (k) a(k) {\cal P}_{\zeta} (k) \; ,
\end{equation}
where $\Delta_l (k)$ is a transfer function. The function $a(k)$ here encodes the possible dependence of SA on the 
scale $k$. It is given by $a(k)=H_0 k^{-1}$ for the LO contribution in sub-scenario A; it should be set to unity in other
cases. Coefficients $B^{LM}_{lm;l'm'}$ are expressed 
in terms of the Wigner 3j-symbols,
\begin{equation}
\nonumber 
B^{LM}_{lm;l'm'}=(-1)^M \sqrt{\frac{(2L+1)(2l+1)(2l'+1)}{4\pi}}
\left (
\begin{array}{ccc} 
L & l & l'\\
0 & 0 & 0
\end{array} 
\right ) \left (
\begin{array}{ccc} 
L & l & l'\\
M & m & -m'
\end{array} 
\right ) \; .
\end{equation}
Finally, ${\bf F}^{ij}$ in Eq.~\eqref{qlm} is the Fisher matrix defined by 
\begin{equation}
\label{fisher} 
F^{ij}_{LM;L'M'} \equiv \langle h^{ij}_{LM} (h^{ij}_{L'M'})^{*} \rangle -\langle h^{ij}_{LM} \rangle 
\langle (h^{ij}_{L'M'})^{*} \rangle  \; .
\end{equation}
The analytic expression for the Fisher matrix in the homogeneous noise
approximation is given by
\begin{equation} 
\label{fishermain}
F^{ij}_{LM;L'M'} =\delta_{LL'} \delta_{MM'} f_{\text{sky}}\sum_{l, l'} 
\frac{(2l+1)(2l'+1)}{16\pi}\left (
\begin{array}{ccc} 
L & l & l'\\
0 & 0 & 0
\end{array} 
\right )^2 \frac{C^2_{l l'} 
\cdot \left(C^{\text{tot},i}_l C^{\text{tot},j}_{l'} +\tilde{C}^{i}_l \tilde{C}^{j}_{l'} \right)}{\left(C^{\text{tot},i}_{l} \right)^2 
\left(C^{\text{tot},j}_{l'}\right)^2} \; .
\end{equation}
Here $C^{\text{tot},i}_l=C_l+N^{i}_l$, where $C_l$ is the standard CMB angular power spectrum and 
$N^{i}_l$ is the angular power spectrum of the homogeneous noise; $\tilde{C}^{i}_l=C^{\text{tot},i}_l$ 
for $i=j$ and $\tilde{C}^{i}_l=C_l$ in the opposite case. We hide the details of the derivation of this 
formula for the case of the single-frequency band analysis in Appendix.

\begin{table}[htb!]
\hspace{1.2cm}
\begin{tabular}{|c|c|c|c|} 
\hline
Model/band &  143 GHz & 217 GHz & $143 \times 217$ GHz\\
\hline 
Sub-scenario A (LO) &$h^2<8.8$ &$h^2 <8.0$ &$h^2< 3.0$ \\
\hline
Sub-scenario A (NLO) & $h^2 \ln \frac{H_0}{\Lambda}<0.34$ & $h^2\ln \frac{H_0}{\Lambda}<0.30$ & $h^2 \ln \frac{H_0}{\Lambda}<0.52$\\
\hline
Sub-scenario B & $h^2<0.0011$ & $h^2<0.0090$ & $h^2 <0.0013$\\
\hline 
Inflation &$|g_*|<0.020$ &$|g_*|<0.020$ & $|g_*|<0.026$\\
\hline 
\end{tabular}
\caption{\footnotesize{Planck 95\% C.L. constraints on the parameter
    $h^2$ of (pseudo)conformal Universe and on the amplitude of the
    axisymmetric quadrupole anisotropy $g_*$.}}\label{Table1}
\end{table}

We use the Planck CMB temperature maps corresponding to the first 15.5 months of
observation  at the frequencies $143$ GHz
and $217$ GHz~\cite{Planck_overview,Planck_maps}. To remove the
contamination of the galactic light and point sources we apply the
High Frequency Instrument (HFI) power spectrum
mask~\cite{Planck_maps, Planck_spec}, which leaves 43\% of the sky
unmasked. Averaging over statistically isotropic realizations is
performed using 100 Planck simulated multi-frequency CMB maps coadded
with the corresponding noise maps and the foreground
maps~\cite{Planck_maps, Planck_iso}. These maps incorporate the effects
of beam asymmetries and complex scanning strategy~\cite{Mitra:2010rt},
which are proved to be crucial for the SI
studies~\cite{Hanson:2010gu,Planck_iso, kim}. The operations with the
maps are performed with the {\it HEALPix} and {\it healpy}
packages~\cite{Gorski:2004by}.

 We implement the estimator~\eqref{qlm} in several steps. First, 
we carry out the inverse-variance filtering using the multigrid
 preconditioner~\cite{Smith:2007rg}; the procedure is discussed in 
detail in Ref.~\cite{Hanson:2009gu}. Second, we evaluate the coefficients
 $C_{ll'}$ using {\it CAMB}~\cite{Lewis:1999bs} and, finally, calculate the sum in
 Eq.~\eqref{hlm} using {\it gsl}~\cite{gsl} and {\it
   slatec}~\cite{slatec} libraries. The range of the multipoles 
studied is set to $2\leq l \leq 1600$. For the larger 
values of $l$, the signal is dominated by the instrumental noise.

In Fig.~\ref{217crosscql} we plot the coefficients 
\begin{equation}
\label{cql}
C^q_L=\frac{1}{2L+1}\sum_M |q_{LM}|^2 \; 
\end{equation}
estimated from the data at frequencies 143 GHz and 217 GHz and their
cross-correlation.  As it is clearly seen, at the level of the
quadrupole the Planck data are in agreement with the hypothesis of
SI. On the other hand, the power of anisotropy contained in $L=4$ of
$143$ and $217$ GHz frequency bands deviates from SI expectations 
at more than $2\sigma$ level. We note that
  the amplitude and the orientation of the $L=4$ multipole are
  different for different frequency bands. Moreover, the significance
  of the excess grows with the increase of the cutoff
  $l_{max}$. Finally, the signal is consistent with the hypothesis of SI in the
  cross-estimator. Therefore we argue that the enhancement at $L=4$
  may originate from statistical fluctuation or systematic effects of
  the noise. The noise dominates at high multipoles and is 
completely uncorrelated between the bands. Consequently the
cross-correlation effectively wipes out all the noise-related
effects. In what follows we consider the cross-estimator as the
preferred method for the study of SA with the Planck data.

We start with constraining the models of the (pseudo)conformal
Universe, where the intermediate stage is absent (sub-scenario A). 
To estimate the
parameter $h^2$, we use the statistics given by the coefficient
$C^q_2$. The remainder of the procedure parallels that employed in
Ref.~\cite{Ramazanov:2013wea}. The final constraints originating from
the LO and SLO contributions to SA are presented in
Table~\ref{Table1}. We note that the limit from the SLO term in
Eq.~\eqref{q2mrolling} is in fact stronger. This is not a surprise,
as the LO term decreases with the wavenumber $k$. Hence, it leaves a
weaker imprint on CMB for the relevant values of the parameter $h$, in
agreement with the findings of previous works~\cite{Ramazanov:2012za,
  Ramazanov:2013wea}.

In Table~\ref{Table1} we also present the constraint on the amplitude
$g_*$ of the SA of the axisymmetric type, ${\cal P}_{\zeta} ({\bf
  k})\propto \left(1+g_* \cos^2 \theta \right)$, where $\theta$ is the
angle between the perturbation wavevector ${\bf k}$ and the direction
of SI breaking. We do this in  view of several inflationary scenarios 
in which SA of this type is generated by vector
fields~\cite{Ackerman:2007nb}. Our limits demonstrate a relatively
mild improvement as compared to ones of Ref.~\cite{kim}, which may be
attributed to the use of the inverse-variance filtering. 
We leave more detailed study of the anisotropic inflationary models 
for our future work~\cite{workin}.

To constrain the parameter $h^2$ in the versions of the (pseudo)conformal
Universe with a long intermediate stage (sub-scenario B), we use the
estimator~\cite{Ramazanov:2012za}
\begin{equation}
\label{estimh2}
h^2 \sum_{L} \frac{(2L+1)F^2_L \tilde{Q}^2_L}{(1+F_L \tilde{Q}_L h^2)^2}=
\sum_L \frac{(2L+1) F_L \tilde{Q}_L}{(1+F_L \tilde{Q}_L h^2)^2}
(F_L C^{q}_L -1) \; ,
\end{equation}
Here $F_L$ are the elements of the Fisher matrix, which we assume to be 
diagonal. 
\begin{figure}[tb!]
\begin{center}
\includegraphics[width=0.22\columnwidth,angle=-90]{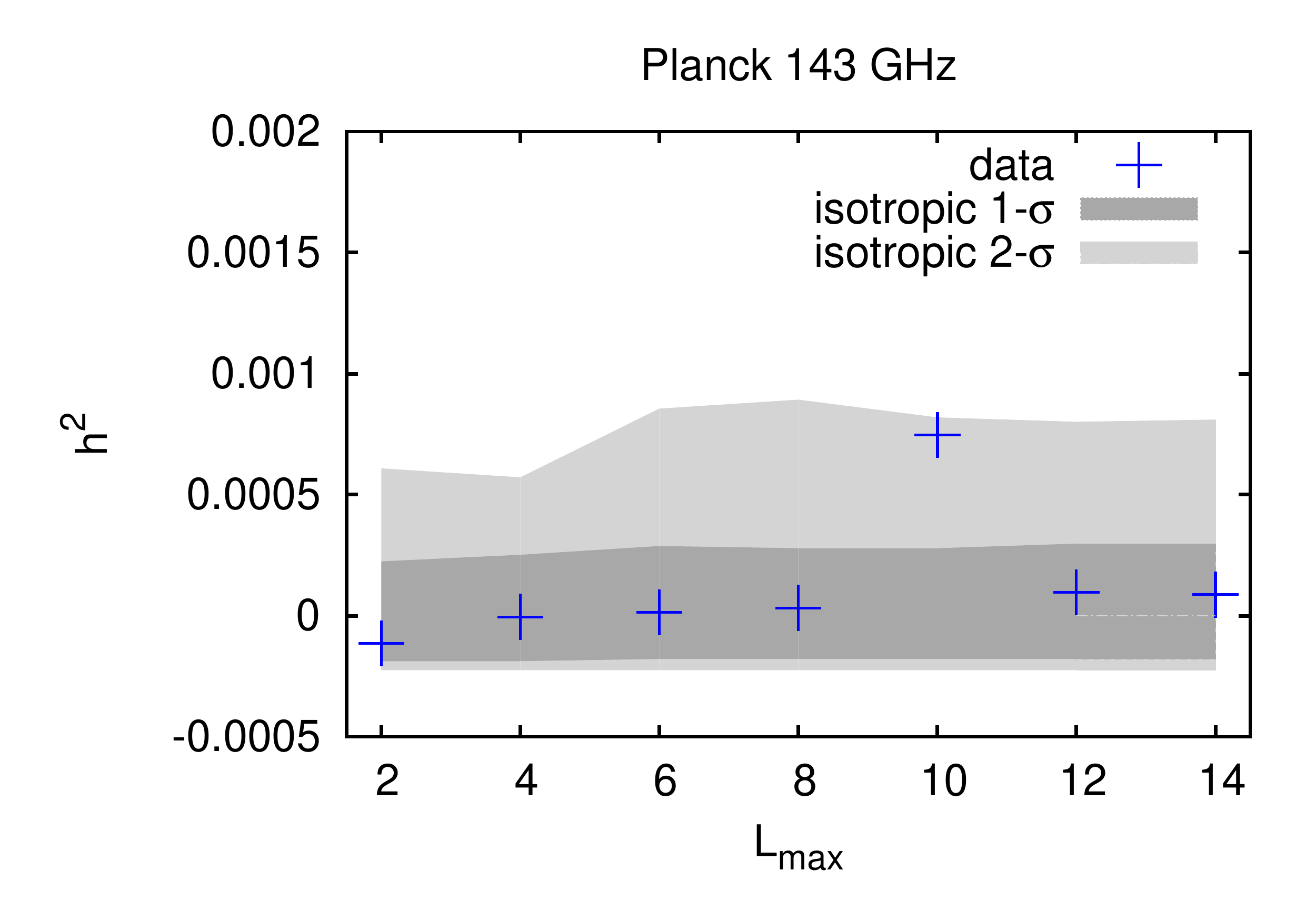}
\includegraphics[width=0.22\columnwidth,angle=-90]{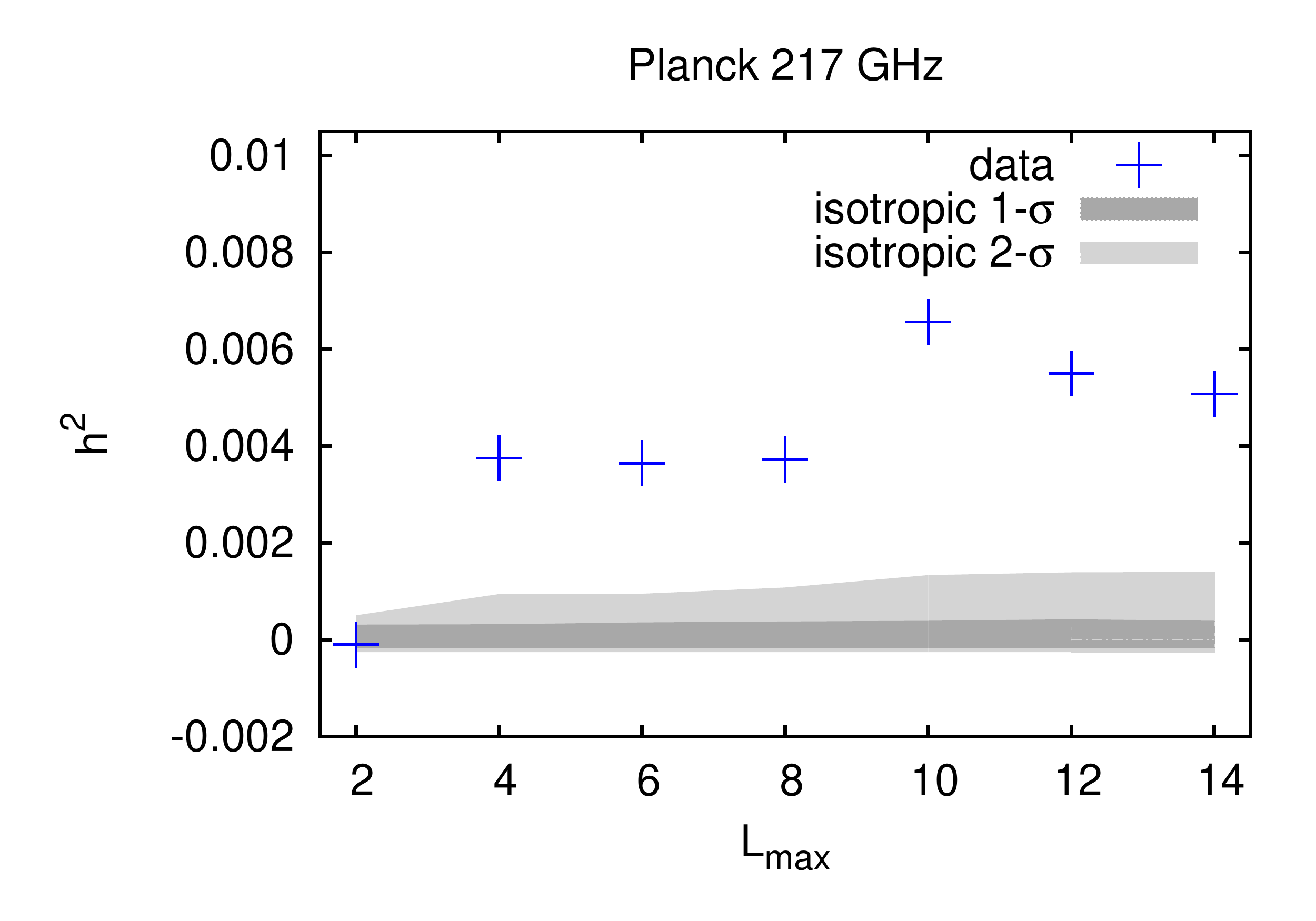}
\includegraphics[width=0.22\columnwidth,angle=-90]{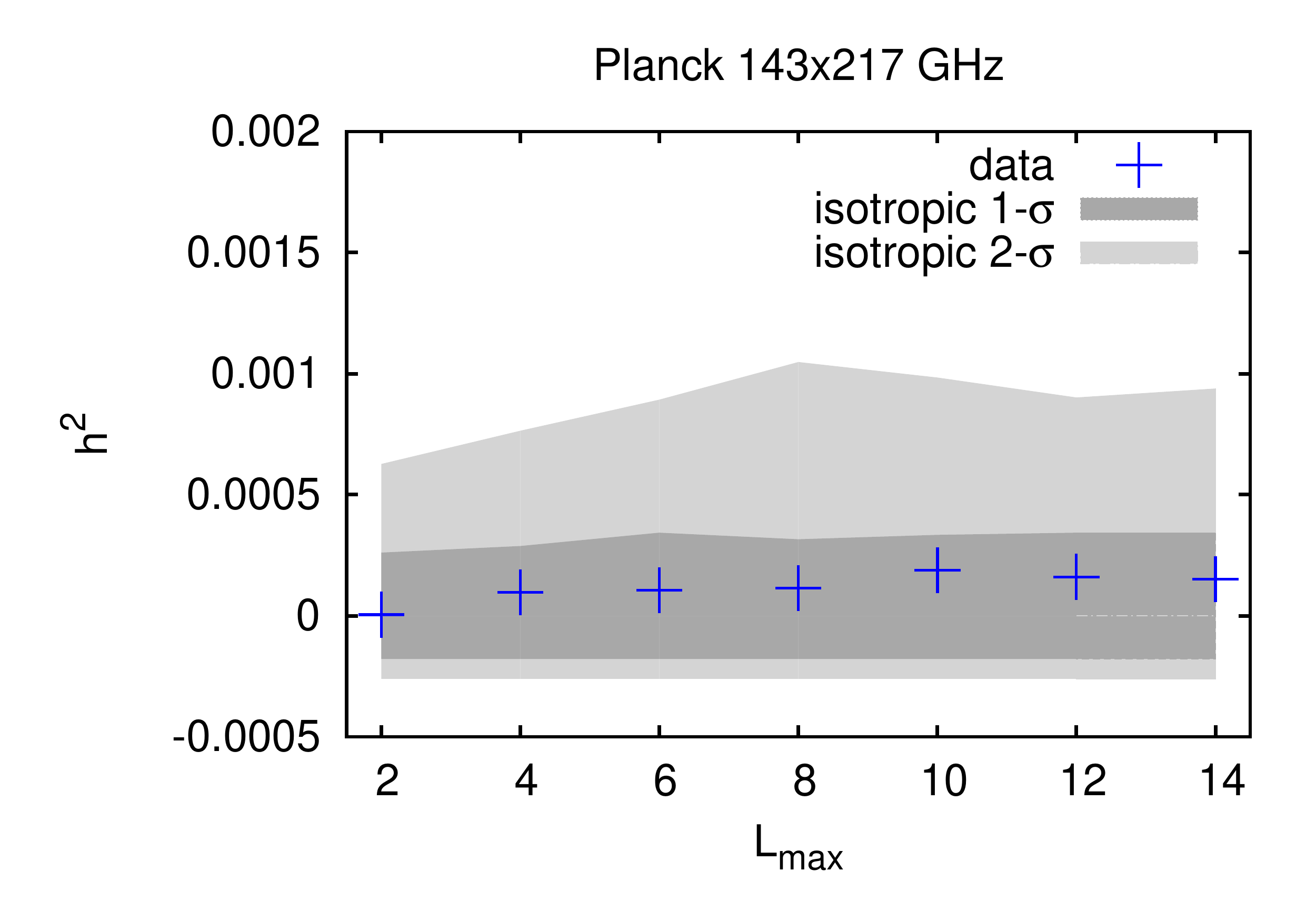}
\end{center}
\caption{Parameter $h^2$ of the (pseudo)conformal Universe with long intermediate stage estimated from the Planck data. 68\% and 95\% C.L. intervals 
are overlaid with dark grey and light grey, respectively.\label{217crossh2}}
\end{figure}
The results of implementing this estimator to the Planck data are
shown in Fig.~\ref{217crossh2}. Values of the estimators are plotted 
for  seven ranges of the multipoles starting from the quadrupole $L=2$ and extending 
up to $L_{max}=2, \dots, 14$. The results for the frequency
band $217$ GHz clearly exhibit SA, which is again due to the enhancement at $L=4$, see Fig.~\ref{217crosscql}. The
rest of the procedure is the same as in the previous
papers~\cite{Ramazanov:2012za, Ramazanov:2013wea}. Final constraints
are presented in Table~\ref{Table1}.

 We conclude that statistical anisotropy is a significant signature 
in the sub-scenario B of the 
(pseudo)conformal Universe, while it is relatively weak in the sub-scenario A. Fortunately, the latter 
may yield strong NG at the trispectrum level~\cite{Libanov:2011bk, Creminelli:2012qr}. The leading contribution to the 
NG in these models occurs in the order $h^2$. Given its mild 
behavior in the folded limit, i.e., 
when two cosmological momenta are anticollinear, it can be compared with the equilateral type NG (rather conservatively). 
Using the existing WMAP limit on the corresponding trispectrum parameter $|\tau^{equil}_{NL}| \lesssim 7 \times 10^{6}$~\cite{Fergusson:2010gn} 
and the estimates presented in~\cite{Libanov:2010nk}, 
one expects to arrive 
at the constraint $h^2 \lesssim 0.1-1$. This is already comparable with the 
constraint deduced from the non-observation of SA. There is in fact another 
contribution to the NG~\cite{Creminelli:2012qr}. Though it emerges in the quartic order in the constant $h$, 
this contribution is enhanced in the folded limit relative to the LO one. 
Remarkably, the SLO NG is precisely of the local type, at least in the folded limit. Making use of the 
Planck 95\% C.L. limit $|\tau^{loc}_{NL}| <2800$~\cite{Ade:2013ydc}, one would expect 
the constraint as strong as $h^2 \lesssim 0.01-0.1$. Hence, NG appears to be 
the most promising signature of
the (pseudo)conformal Universe without the intermediate stage. 
The detailed analysis of the trispectrum remains to be performed, 
however.

{\it Acknowledgments.} We thank V.~Rubakov, M.~Thorsrud and F.~Urban
for numerous fruitful discussions. We are grateful to D. Hanson for
kindly providing the code for inverse-variance filtering.  This work
is supported by the Russian Science Foundation grant 14-12-01430
(G.R.)  and Belgian Science Policy IAP VII/37 (S.R.).
G.R. acknowledges the fellowship of the Dynasty foundation. The
numerical part of the work was done at the cluster of the Theoretical
Division of INR RAS.

\section*{Appendix}
In this Appendix, we provide the analytical computation of the 
Fisher matrix in the homogeneous noise approximation for the realistic case of the masked 
sky. We do this for the single-frequency band analysis. The generalization to the 
multi-frequency-band analysis is straightforward. We start with
substituting Eq.~\eqref{hlm} into Eq.~\eqref{fisher} and obtain 
\begin{equation}
\label{fij}
\begin{split}
F_{LM;L'M'} =(-1)^{M'}\frac{1}{4} \sum_{ll';mm'} \sum_{\tilde{l} \tilde{l}'; \tilde{m} \tilde{m}'} i^{l'-l+\tilde{l}-\tilde{l}'} C_{ll'} C_{\tilde{l} \tilde{l}'} 
B^{LM}_{lm;l'm'} B^{L',-M'}_{\tilde{l} \tilde{l}'; \tilde{m} \tilde{m}'} \\ \Bigl( \langle \bar{a}_{l,-m} \bar{a}_{\tilde{l}, -\tilde{m}} \rangle \langle \bar{a}_{l'm'} \bar{a}_{\tilde{l}', \tilde{m}'} \rangle+ 
\langle \bar{a}_{l,-m} \bar{a}_{\tilde{l}', \tilde{m}'} \rangle \langle \bar{a}_{\tilde{l}, -\tilde{m}}  \bar{a}_{l'm'} \rangle \Bigr) \; .
\end{split}
\end{equation}
Here we made use of the Isserlis-Wick theorem. The relation between the 
coefficients $\hat{a}_{lm}$ and spectral coefficients obtained from the 
hypothetical full sky analysis $\hat{a}^{f}_{lm}$ is given by~\cite{Bartolo:2004if}
\begin{equation}\nonumber
\hat{a}_{lm}= \sum_{l'm'} W_{lm;l'm'} \hat{a}^{f}_{l'm'} \; .
\end{equation}
Here $W_{lm;l'm'}$ is the transition matrix
\begin{equation}
\label{wlm}
W_{lm;l'm'} =\int d{\bf n} W({\bf n}) Y^*_{lm} ({\bf n}) Y_{l'm'} ({\bf n}') \; .
\end{equation}
$W({\bf n})$ is the ``mask'' function, which takes in the case of the
sharp mask the value $1$ in the unmasked pixels and zero otherwise. 
Then the unmasked fraction of the sky $f_{sky}$ is given by the
integral of $W({\bf n})$ over the sphere,
\begin{equation}
\label{fsky}
f_{sky}=\int \frac{d{\bf n}}{4\pi} \cdot W({\bf n})  \; .
\end{equation}
Neglecting the commutator between the masking and inverse-variance
filtering procedures, one arrives to the following relation for the
filtered harmonic coefficients
\begin{equation}
\label{almtrue}
\bar{a}_{lm}= \sum_{l'm'} W_{lm;l'm'} \bar{a}^{f}_{l'm'} \; .
\end{equation}

Substituting Eq.~\eqref{almtrue} into Eq.~\eqref{fij}, we obtain 
\begin{equation}
\label{fishinter}
\begin{split} 
F_{LM;L'M'} &=(-1)^{M'}\frac{1}{2} \sum_{ll';mm'} \sum_{\tilde{l} \tilde{l}';\tilde{m} \tilde{m}'} i^{l'-l+\tilde{l} -\tilde{l}'}C_{ll'} C_{\tilde{l} \tilde{l}'}
\Bigl(C^{tot}_l C^{tot}_{\tilde{l}}C^{tot}_{l'} C^{tot}_{\tilde{l}'} \Bigr)^{-1} B^{LM}_{lm;l'm'} B^{L',-M'}_{\tilde{l}\tilde{l}';\tilde{m} \tilde{m}'} \times \\ 
& \times \sum_{n=0} \sum^{n}_{k=-n} \sum_{n'=0} \sum^{n'}_{k'=-n'} (-1)^{k+k'}C^{tot}_n C^{tot}_{n'} 
 W_{ln;-m,k} W_{\tilde{l} n; -\tilde{m},-k} W_{l'n';m'k'} W_{\tilde{l}'\tilde{n}';\tilde{m}',-\tilde{k}'} \; .
\end{split}
\end{equation}

Here we made use of the following relations, which are valid in the homogeneous noise approximation
\begin{equation}
\nonumber 
\bar{a}^{f}_{lm}= (C^{tot}_l)^{-1} \hat{a}^{f}_{lm} \; , 
\end{equation}
and
\begin{equation}
\nonumber 
\langle \hat{a}^{f}_{lm} \hat{a}^{f}_{l'm'} \rangle =(-1)^{m}C^{tot}_l \delta_{ll'} \delta_{m,-m'} \; .
\end{equation}
Replacing sufficiently slowly varying functions $C^{tot}_n$ and $C^{tot}_{n'}$ by $C^{tot}_l$ and $C^{tot}_{l'}$, we obtain
\begin{equation}
\label{flminter}
\begin{split}
F_{LM;L'M'} & \approx \frac{1}{2} \sum_{ll';mm'} \sum_{\tilde{l} \tilde{l}';\tilde{m} \tilde{m}'} i^{l'-l+\tilde{l} -\tilde{l}'} (-1)^{M+m+\tilde{m} +m'+\tilde{m}'} C_{ll'} C_{\tilde{l} \tilde{l}'} 
\Bigl(C^{tot}_{\tilde{l}} C^{tot}_{\tilde{l}'} \Bigr)^{-1} \times \\ 
& \times \int d{\bf n} Y_{lm} ({\bf n}) Y_{l',-m'} ({\bf n}) Y_{LM} ({\bf n}) 
\int d{\bf \tilde{n}} Y_{\tilde{l} \tilde{m}}({\bf \tilde{n}}) Y_{\tilde{l}',-\tilde{m}'} ({\bf \tilde{n}}) Y_{L',-M'} ({\bf \tilde{n}}) \times \\
& \times \int d {\bf n}_1 W({\bf n}_1) Y_{lm} ({\bf n}_1) Y_{\tilde{l}, \tilde{m}} ({\bf n}_1) 
\int d{\bf n}_2 W({\bf n}_2) Y_{l',-m'} ({\bf n}_2) Y_{\tilde{l}', -\tilde{m}'} ({\bf n}_2) \; .
\end{split}
\end{equation}
Let us comment on the derivation of this formula. 
First, we replaced the coefficients $W_{lm;l'm'}$ in Eq.~\eqref{fishinter} by Eq.~\eqref{wlm} and coefficients $B^{LM}_{lm;l'm'}$ by the integral over spherical harmonics
\begin{equation}
\nonumber 
B^{LM}_{lm;l'm'} =\int d {\bf n} Y^{*}_{lm} ({\bf n}) Y_{l'm'} ({\bf n}) Y_{LM} ({\bf n}) \; ,
\end{equation}
respectively. We then provided the summation over the indexes $n$, $k$, $n'$ and $k'$. At this 
point the following relation is used
\begin{equation}
\label{sumharm}
\sum_{l=0} \sum^{l}_{m=-l} Y^{*}_{lm} ({\bf n})Y_{lm} ({\bf n}')=\delta ({\bf n}-{\bf n}')
\end{equation}
The expression~\eqref{flminter} is obtained by integrating out the
delta-functions. Next, we again replace slowly changing functions
$C_{\tilde l}$ and $C_{\tilde {l'}}$ by $C_l$ and $C_{l'}$,
respectively and sum over $(\tilde{l}, \tilde{m})$ and $(\tilde{l}',
\tilde{m}')$ in Eq.~\eqref{flminter} using Eq.~\eqref{sumharm}. We arrive at
\begin{equation}
\begin{split}
F_{LM;L'M'} &\approx \frac{1}{2} \sum_{ll';mm'} \frac{C^2_{ll'}}{C^{tot}_l C^{tot}_{l'}} \int d{\bf n} Y_{l,-m} ({\bf n}) Y_{l'm'} ({\bf n}) Y^{*}_{L,-M} ({\bf n}) \times \\ 
& \times \int d {\bf n}' W({\bf n}') Y^{*}_{l,-m} ({\bf n}') Y^{*}_{l'm'} ({\bf n}') Y_{L',-M'} ({\bf n}') \; .
\end{split}
\end{equation}
Finally, we sum over $m$ and $m'$ indexes, 
\begin{equation}
\label{fish}
\begin{split}
F_{LM;L'M'} & \approx \frac{1}{2} \sum_{ll'} \left(\frac{2l+1}{4\pi} \right) 
\left( \frac{2l'+1}{4\pi} \right)\frac{C^2_{ll'}}{C^{tot}_l C^{tot}_{l'}} \times \\ 
&\times \int d{\bf n} d{\bf n}'  W({\bf n}') P_l ({\bf n}{\bf n}') P_{l'} ({\bf n}{\bf n}') Y^{*}_{L,-M} ({\bf n}) Y_{L',-M'} ({\bf n}') \; .
\end{split}
\end{equation}
The expression~\eqref{fish} may be simplified by using the fact that the matrix~\eqref{fish} is approximately 
diagonal, and has a sufficiently mild dependence on the numbers $M$
and $M'$. The Fisher matrix may be therefore approximated as
\begin{equation}
\nonumber 
F_{LM;L'M'} \approx F_L \delta_{LL'} \delta_{MM'} \; ,
\end{equation}
where
\begin{equation}
F_L \approx \frac{1}{2L+1} \sum_M F_{LM;LM} \; . 
\end{equation}
Taking the sum over $M$ index we arrive at 
\begin{equation}
\nonumber 
F_L \approx \frac{1}{8\pi} \sum_{ll'} \left(\frac{2l+1}{4\pi} \right) 
\left( \frac{2l'+1}{4\pi} \right)\frac{C^2_{ll'}}{ C^{tot}_l C^{tot}_{l'}} \int d{\bf n} d{\bf n}'  W({\bf n}') P_l ({\bf n}{\bf n}') P_{l'} ({\bf n}{\bf n}') P_{L} ({\bf n} {\bf n}') \; .
\end{equation}
The integral over ${\bf n}$ may be taken,
\begin{equation}
\nonumber 
F_{L} \approx \sum_{ll'} \frac{(2l+1)(2l'+1)}{8\pi}\frac{C^2_{ll'}}{C^{tot}_l C^{tot}_{l'}} \left (
\begin{array}{ccc} 
L & l & l'\\
0 & 0 & 0
\end{array} 
\right )^2\int \frac{d{\bf n}'}{4\pi} \cdot W({\bf n}') 
\end{equation}
Taking into account Eq.~\eqref{fsky}, we arrive at
Eq.~\eqref{fishermain} of the main text of the paper. In particular,
the calculation justifies the presence of the factor $f_{sky}$ in the
approximate Fisher matrix.

\end{document}